\documentclass[mathleft
]{an}
\usepackage{epsfig,graphicx}
\usepackage{times}
\usepackage{natbib}
\newcommand{\solphys}{{Sol. Phys.}}
\newcommand{\aap}{{A\&A}}
\newcommand{\nat}{{Nature}}
\begin{document}

\Pagespan{1}{}
\Yearpublication{}%
\Yearsubmission{2010}%
\Month{}%
\Volume{}%
\Issue{}%

\title{Can major solar flares excite high-frequency global waves in the Sun ?}

\author{Brajesh Kumar\inst{1}\fnmsep\thanks{Corresponding author:
  \email{brajesh@prl.res.in}\newline}
\and  Savita Mathur\inst{2,3}
\and  R. A. Garcia\inst{4}
\and  P. Venkatakrishnan\inst{1}
}
\titlerunning{Flare excited global waves in the Sun}
\authorrunning{B. Kumar, S. Mathur, R. A. Garc\'ia \& P. Venkatakrishnan}
\institute{
Udaipur Solar Observatory, Physical Research Laboratory, Dewali, Badi Road, Udaipur 313 004, India
\and
Indian Institute of Astrophysics, Koramangala, Bangalore 560 034, India
\and
High Altitude Observatory, 3080, Center Green Drive, Boulder, CO 80302, USA
\and
Laboratoire AIM, CEA/DSM-CNRS, Universit\'e Paris 7 Diderot, IRFU/SAp, Center de Saclay, 91191, Gif-sur-Yvette, France}

\received{}
\accepted{}
\publonline{later}


\abstract{
The study of low-degree high-frequency waves in the Sun can provide new insight into the 
dynamics of the deeper layers of the Sun. Here, we present the analysis of the velocity observations
of the Sun obtained from the Michelson and Doppler Imager (MDI) and Global Oscillations at Low
Frequency (GOLF) instruments on board Solar and Heliospheric Observatory ({\em SOHO}) spacecraft for
the major flare event of 2003 October 28 during the solar cycle 23. We have applied wavelet transform 
to the time series
of disk-integrated velocity signals from the solar surface using the full-disk Dopplergrams obtained
from MDI. The wavelet power spectrum computed from MDI velocity series clearly shows that there 
is enhancement of high-frequency global waves in the Sun during
the flare. We do observe this signature of flare in the Fourier Power Spectrum of these
velocity oscillations. However, the analysis of disk-integrated velocity observations
obtained from GOLF shows only feeble effect of flare on high-frequency oscillations.}

\maketitle

\section{Introduction}

\sloppy
The normal modes of oscillations of the Sun peak in the frequency regime 2-4 mHz and are known as
$p$ modes. Apart from the normal $p$ modes, researchers have found the presence of
high-frequency oscillations
(frequencies higher than the solar-photospheric acoustic cut off at $\sim$ 5.3 mHz) in the solar-acoustic
spectrum \citep{libb88a, libb88b, garcia98, chap03, jim05, karoff08}. Unlike
the $p$ modes, the driving force for these high-frequency solar oscillations is still not clearly 
understood. On one hand, \cite{balm90} suggest that the high-frequency waves are partly reflected by the
sudden change in temperature at the transition region between the chromosphere and the corona, while 
\cite{kumar91} explain these high-frequency waves as an interference phenomenon between ingoing
and outgoing waves from a localized source just beneath the photosphere.

\sloppy
Just after the advent of helioseismology in 1970s, \cite{wolff72} suggested that large
solar flares can stimulate free modes of oscillation of the
entire Sun, by causing a thermal expansion that would drive a compression front to move
into the solar interior. First of all, \cite{haber88} reported
an average increase in the power of intermediate-degree modes after a major flare
(of class X13/3B) using a few hours of solar-oscillations data. However, 
\cite{braun90} could not detect acoustic-wave excitation
from an X-class flare. \cite{koso98} reported the first detection
of ``solar quakes'' inside the Sun, caused by the X2.6 flare of 1996 July 9.
Following this result, \cite{donea99}
found an acoustic source associated with a flare using seismic images produced with
helioseismic-holography technique. Application of ring-diagram analysis showed that the
power of the global $p$ modes appears
to be larger in several flare-producing active regions as compared with the power in
non-flaring regions of similar magnetic field strength \citep{ambastha03}.

Further, \cite{donea05} have reported emission of seismic waves from large
solar flares using helioseismic holography. Some of the large solar flares have been observed to
produce enhanced high-frequency acoustic velocity oscillations in localized parts of active regions
\citep{kumar06}. \cite{venkat08} observed co-spatial evolution of seismic sources and
H-alpha flare kernels and large downflows in the seismic sources during the flare event 
of 28 October 2003. A search for a correlation between
the energy of the low-degree $p$ modes and flares using velocity observations
of the Sun remained inconclusive \citep{foglizzo98, gavry99, chap04, ambastha06}. The study of
low-degree high-frequency (LDHF) waves in the Sun is important as this can bring more constraints
on the rotation profile between 0.1 and 0.2~$R_\odot$ \citep{garcia_new08, mathur08}. The 
study of the effect of flares
on such LDHF waves can provide a clue for the origin of these waves.

Recently, \cite{karoff08} have reported that the correlation between X-ray flare
intensity and the energy in the acoustic spectrum of disk-integrated intensity oscillations
(as observed with VIRGO (Variability of Solar IRradiance and Gravity) (\cite{frohlich95})
instrument on board {\em SOHO}) is stronger for
high-frequency waves than for the well known 5-minute oscillations. In this
study, we have searched
for the effects of flares in the time series of disk-integrated velocity
signals from the solar surface using the full-disk Dopplergrams obtained from the MDI instrument.
We have also looked for these effects in disk-integrated velocity observations obtained from
the GOLF (Global Oscillation at Low Frequency)
\citep{gabriel95, gabriel97} instrument on board {\em SOHO}. These studies have been applied to the
major solar flare of 2003 October 28 (of class X17.6/4B) that occurred in the solar cycle 23.
Wavelet and Fourier analyses of MDI
velocity observations clearly indicate the enhancement in high-frequency global waves in the
Sun during the flare. However, this signature of flare is weaker in the case of GOLF
as compared to MDI data.

\section{DATA ANALYSIS AND RESULTS}

\begin{figure}
\centering
\includegraphics[width=0.23\textwidth]{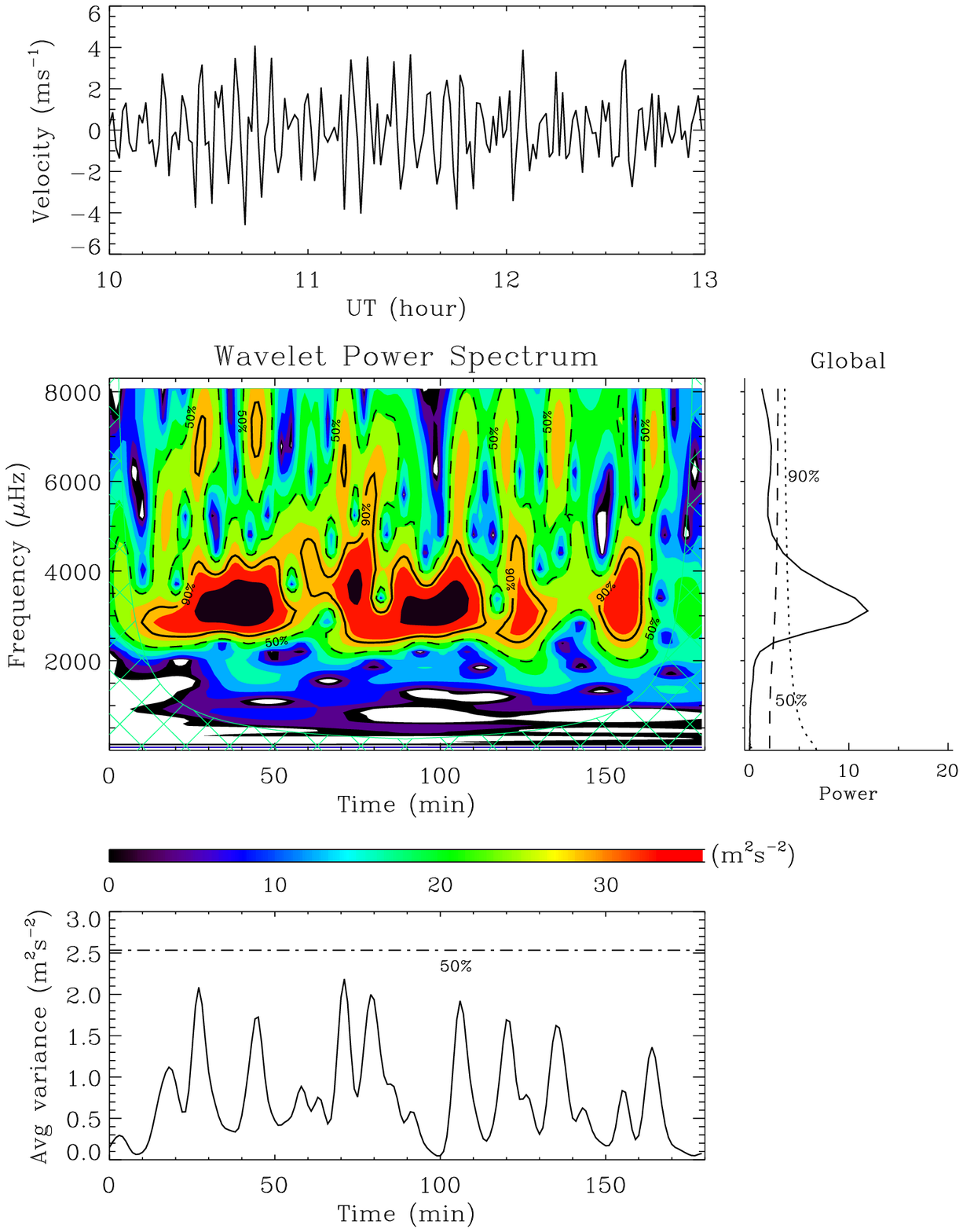} \hspace*{0.25 cm}
\includegraphics[width=0.23\textwidth]{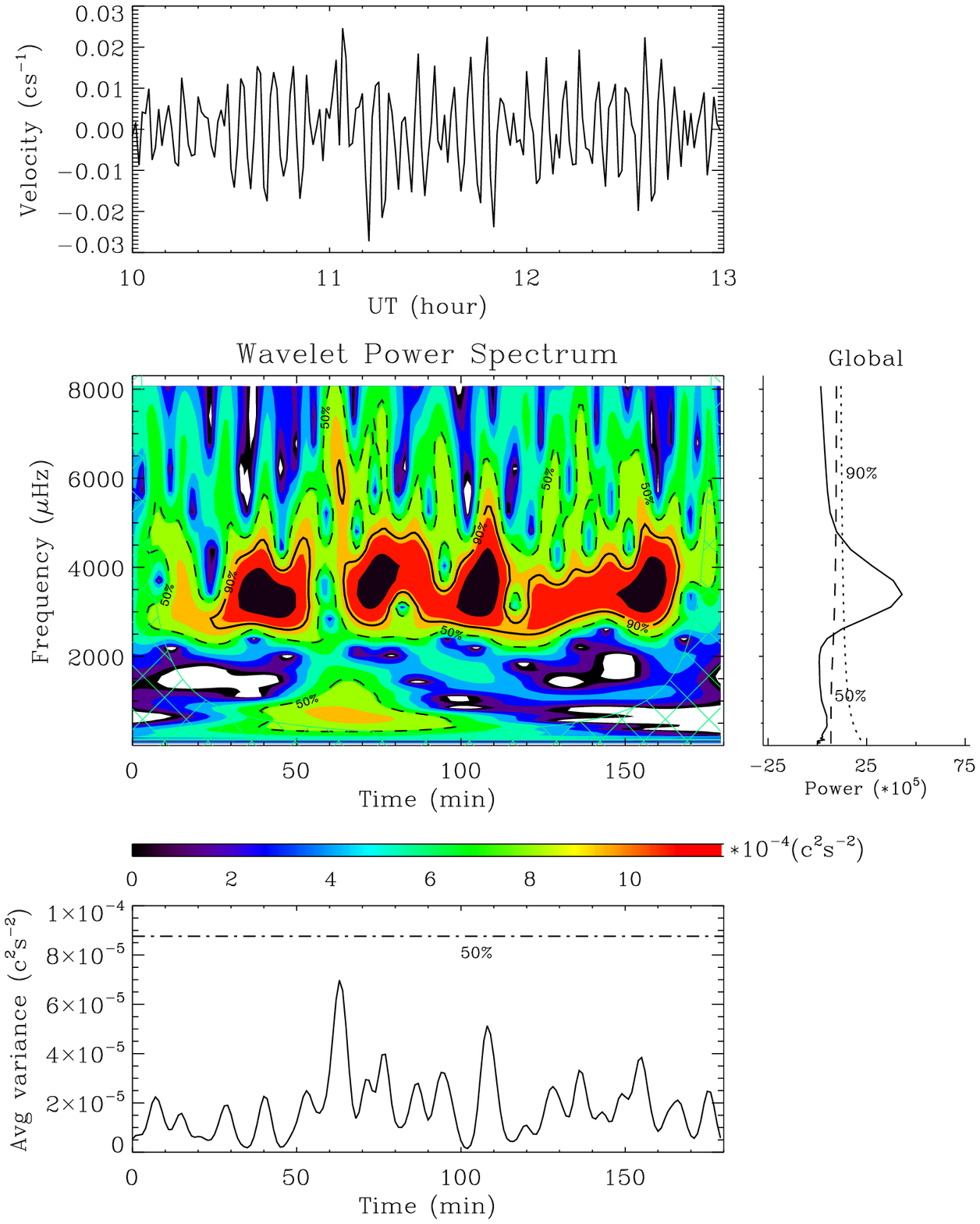}\\
\hspace*{0.4cm}(a)
\hspace*{4.0cm}(b)
\caption{(a) The upper panel shows the temporal evolution of disk-integrated velocity signals obtained from the
full-disk Dopplergrams observed by MDI during 10:00-13:00~UT spanning the flare event of 2003 October 28.
The middle panels show the WPS and the GWPS (see text for details). The bottom panel illustrates the scale-average
time series for the WPS in the frequency regime 5-8~mHz. The dashed-dotted line corresponds to 50\%
significance level of the average variance. (b) Same as above, but using GOLF data.}
\end{figure}

\begin{figure}
\centering
\includegraphics[width=0.23\textwidth]{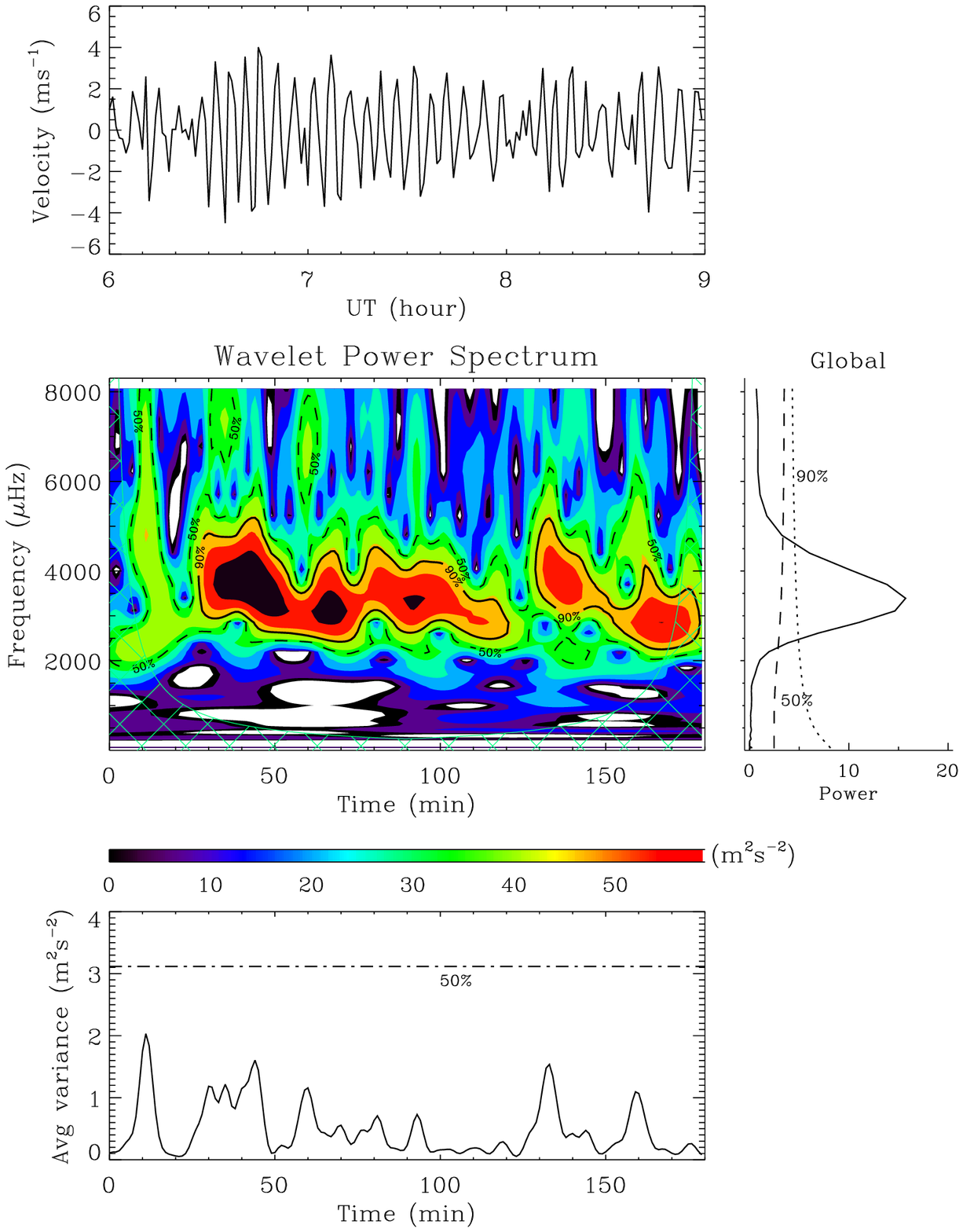} \hspace*{0.25 cm}
\includegraphics[width=0.23\textwidth]{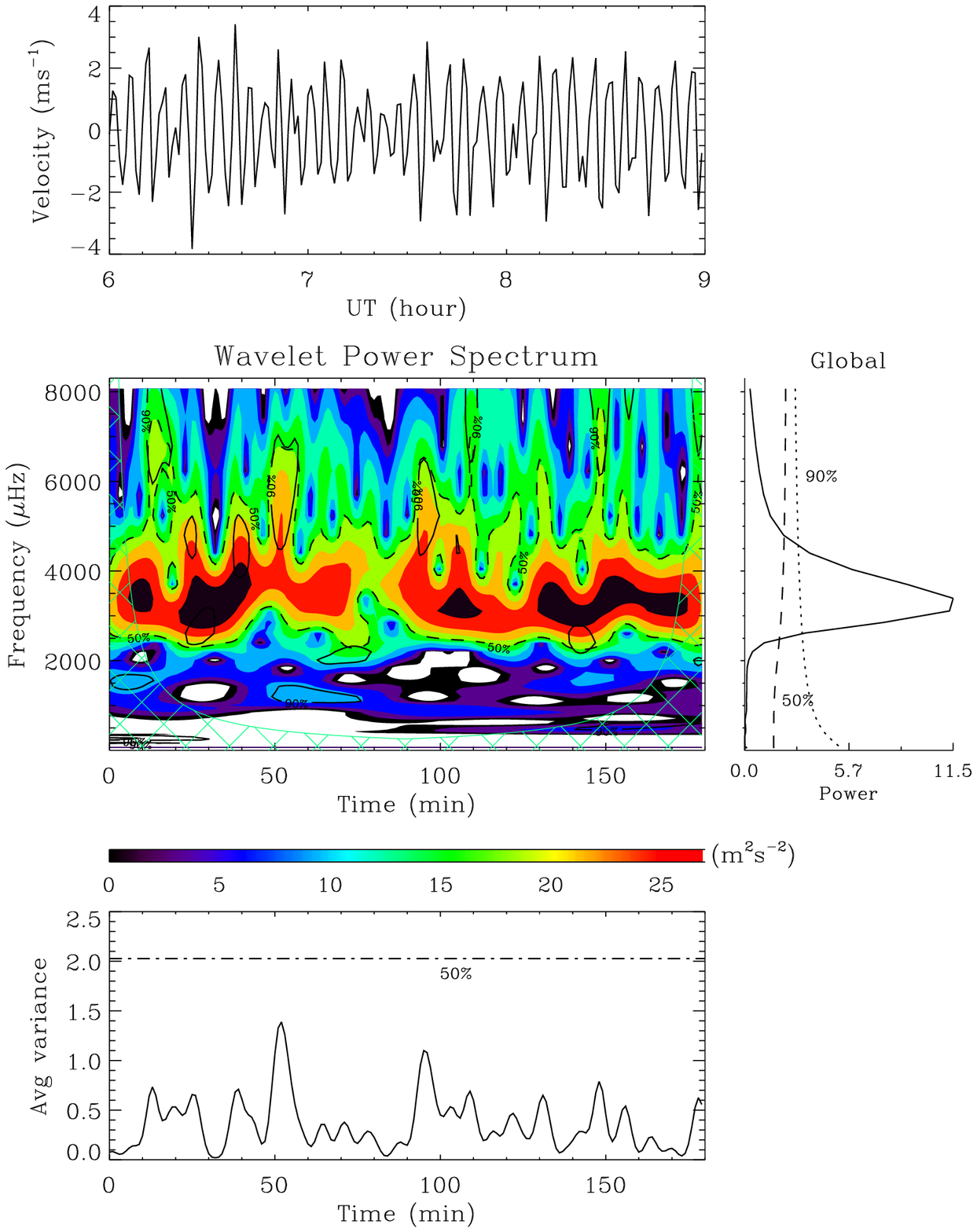}\\
\hspace*{0.4cm}(a)
\hspace*{4.0cm}(b)
\caption{Same as Figure~1, but for a quiet period (non-flaring condition) using (a) MDI data
and (b) GOLF data.}
\end{figure}

\begin{figure}
\centering
\includegraphics[width=0.16\textwidth, angle=90]{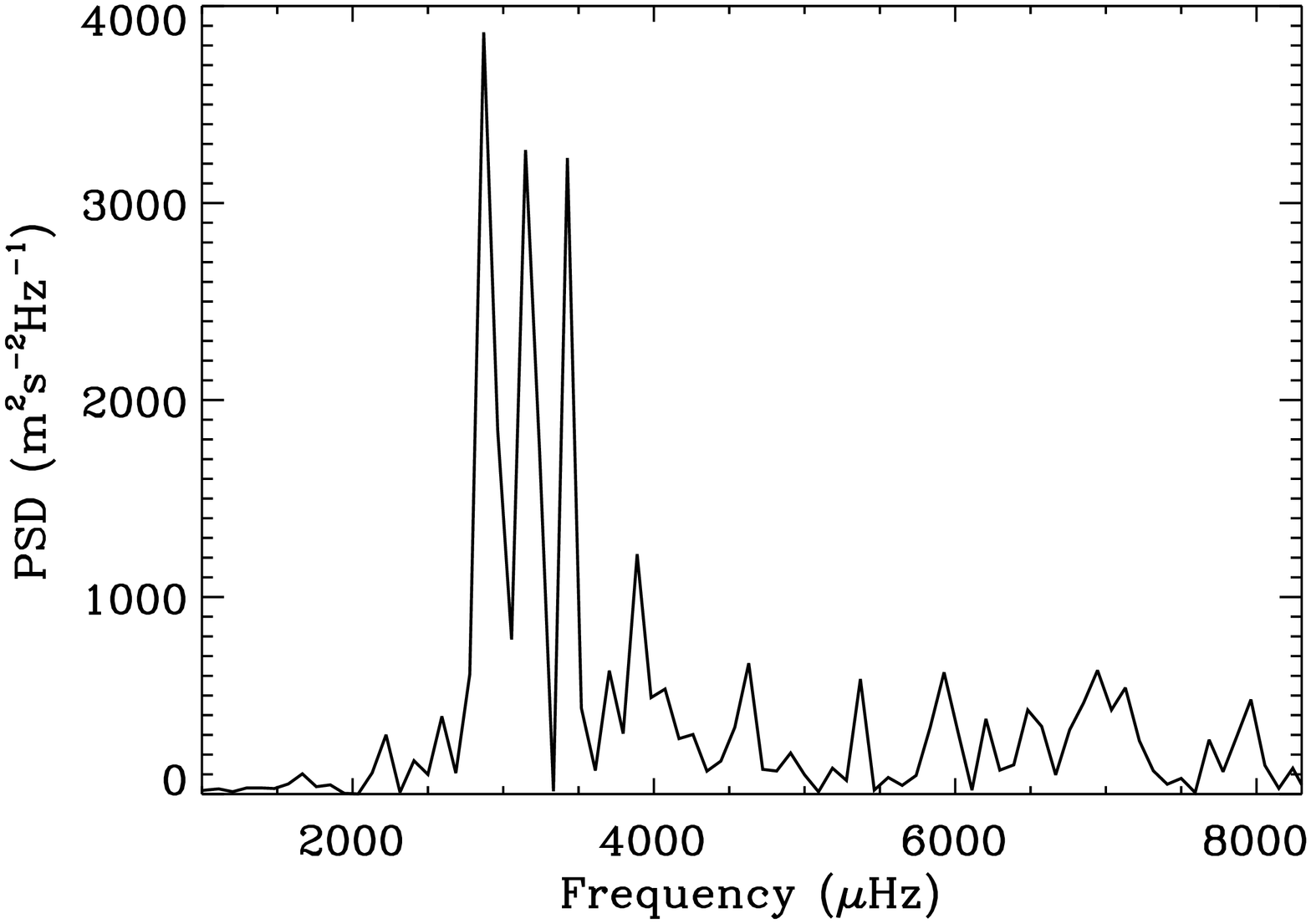} \hspace*{0.25 cm}
\includegraphics[width=0.16\textwidth, angle=90]{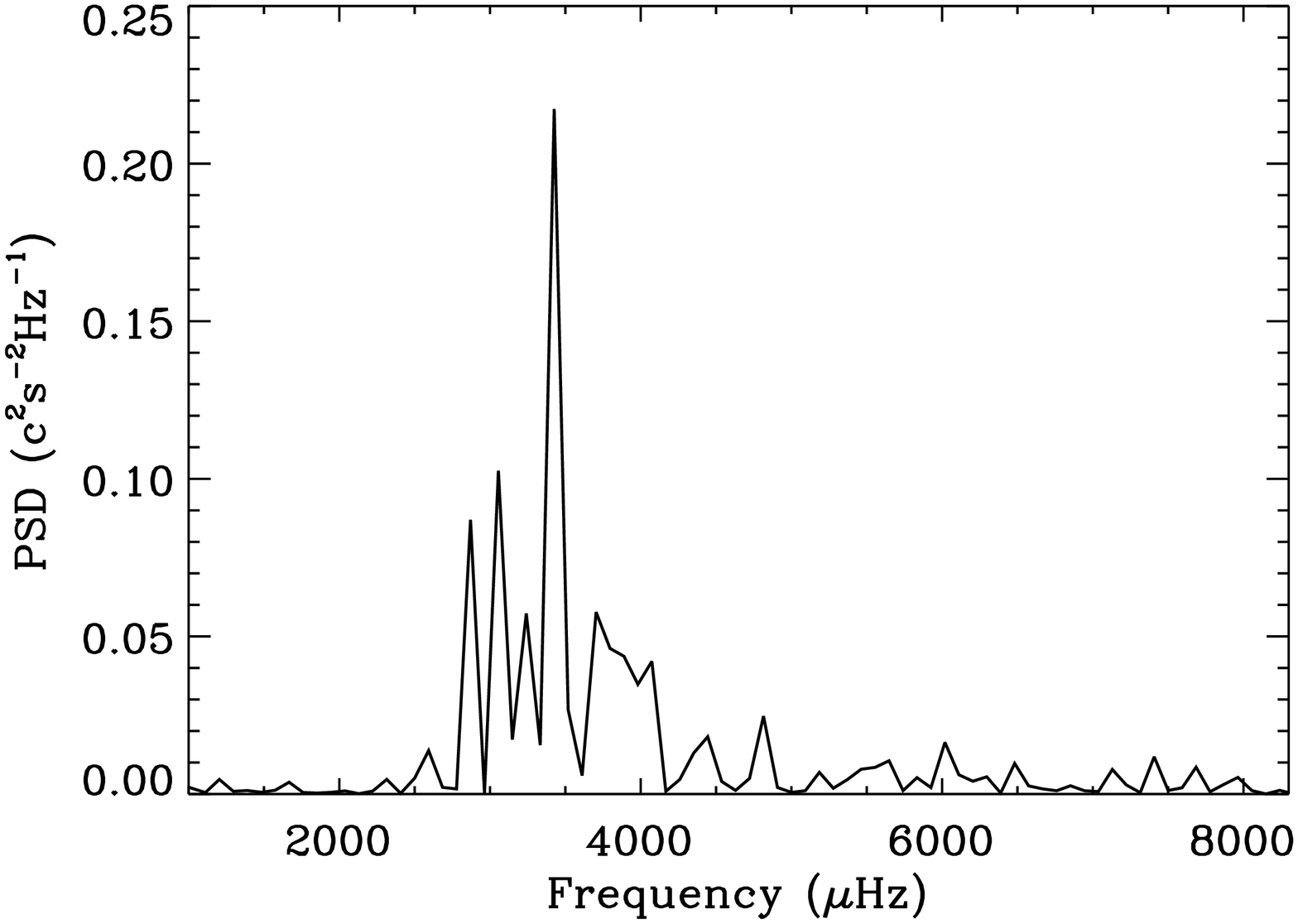}\\
\vspace*{0.25 cm}
\includegraphics[width=0.16\textwidth, angle=90]{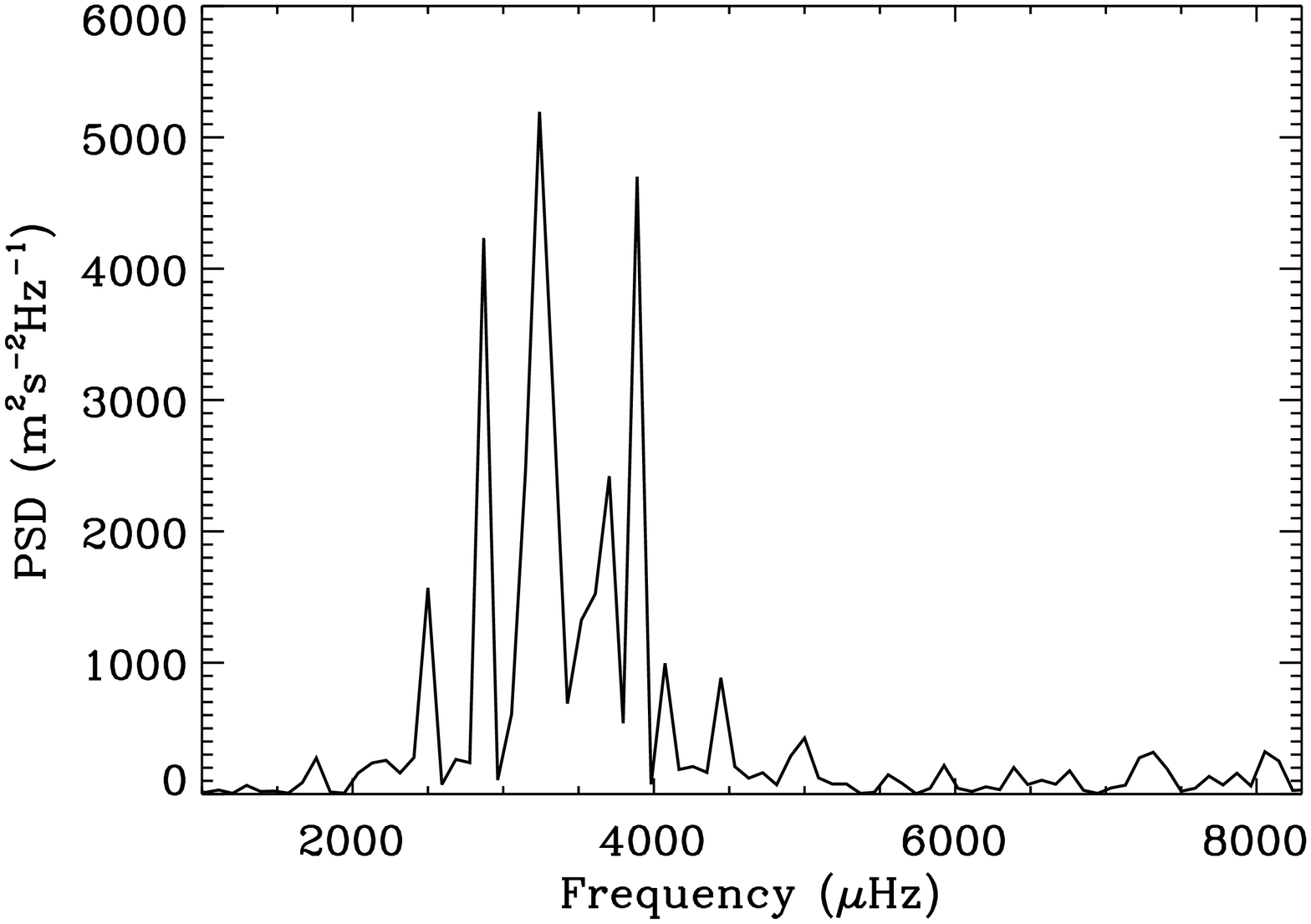} \hspace*{0.25 cm}
\includegraphics[width=0.16\textwidth, angle=90]{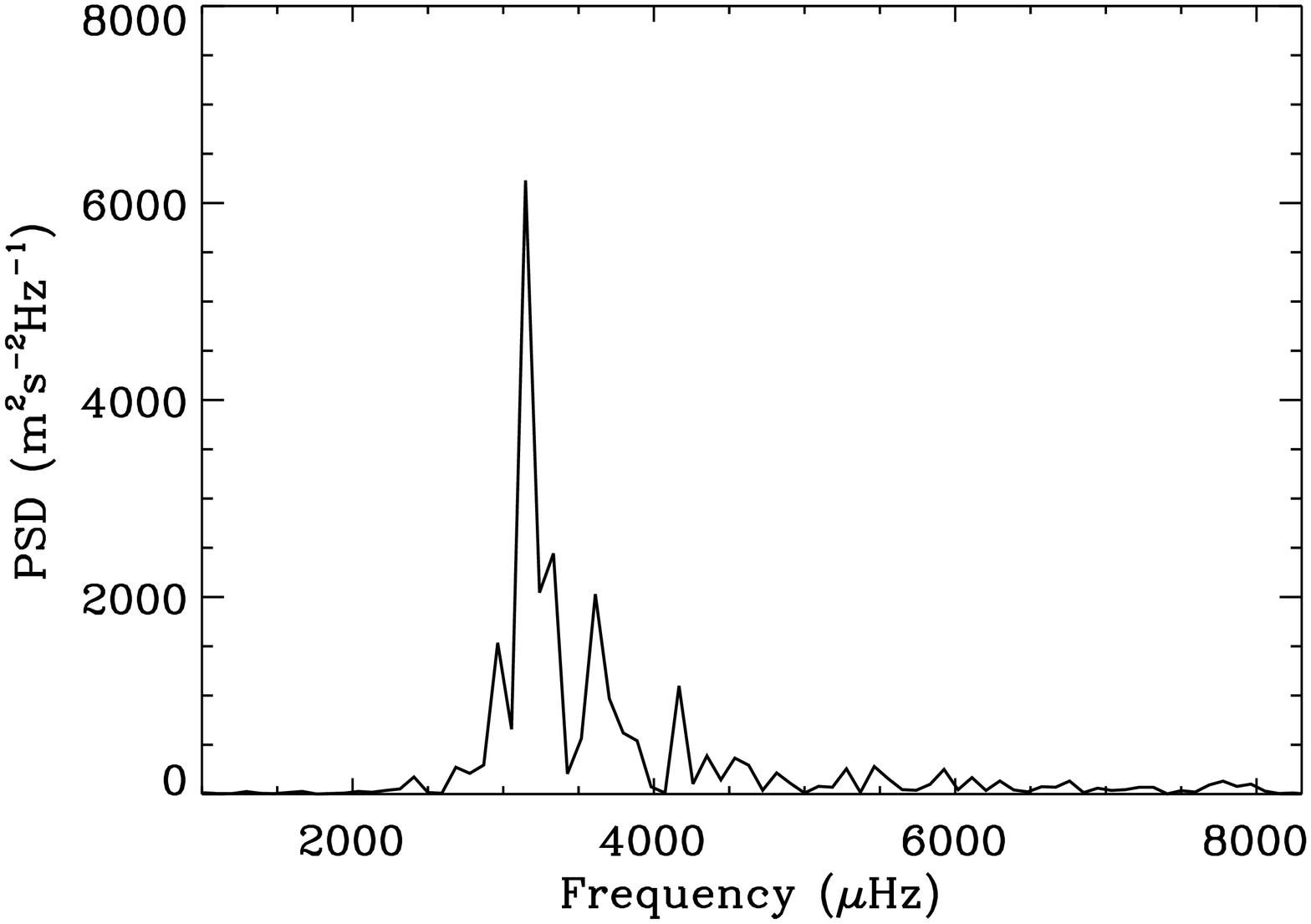}\\
\caption{Fourier Power Spectrum (FPS) estimated from velocity time series. The left panels 
illustrate the FPS
obtained from the three hours of the MDI data for the flare event of 2003 October 28 
during 10:00-13:00~UT (top left panel), and a quiet period (non-flaring condition) 
(bottom left panel), respectively. The right panels are obtained from the corresponding GOLF data.}
\end{figure}

\subsection{DATA USED FROM MDI}

In normal mode of observation, MDI obtains full-disk Dopplergrams of the Sun at a cadence of 60~s and 
spatial sampling rate of 2 arcsec per pixel. We have used a sequence of such Dopplergrams for 
three hours (10:00--13:00~UT) spanning the flare obtained on 2003 October 28.
We have also used three hours of MDI
Dopplergrams for a quiet period (non-flaring condition) as control data. In order to compare the
temporal behaviour with the
disk-integrated intensity observations by VIRGO, we have summed up the velocity signals over all
the pixels of the MDI full-disk Dopplergrams. A two-point
backward difference filter (difference between two consecutive measurements)
is applied to these sequences of images to enhance the velocity signals
from the $p$ modes and high-frequency waves above the other solar background features. 
The sequence of these filtered Doppler
images are then collapsed into a single velocity value, excluding
the noisy pixels along the solar limb. This process is applied to the time series of Doppler images
for the aforementioned flare event as well as the quiet period. It is believed that by
collapsing the
full-disk Doppler images, the acoustic modes with maximum $l$=0,1,2,3 remain while the modes higher than
these are averaged out. Thus, the collapsed velocity value should be the representative of the global
acoustic modes. The temporal evolution of these disk-integrated velocity signals for three hours
spanning the flare is shown in the upper panel of Figure~1(a) and that for the
quiet period is shown in Figure~2(a).

\subsection{DATA USED FROM GOLF}

The GOLF instrument measures the disk-integrated line-of-sight velocity
of the Sun at a cadence of one raw count every 10~s. However, we have rebinned the velocity data
from GOLF for every 60 s to match with MDI. We have used the same periods of data for both instruments
for the flare event of 2003 October 28, as well as for the non-flaring condition. We have
worked with the standard velocity time
series \citep{garcia05, ulrich00} as well as the original raw-counting rates. We have also checked if
there was any
anomalous behavior during the analyzed periods of time in the housekeeping data. Indeed, during the
flare event of 2003 October 28, GOLF raw-counting rates of both photomultipliers suffered an increase
in the measurements probably as a consequence of the impact of high-energetic particles. Due to these
contaminations, the standard velocity time series have been filtered out during this period and thus,
we have been obliged to work with the raw observations obtained by GOLF during the aforementioned
flare event. A two-point backward difference
filter is applied to the velocity series to remove the effect of the rotation and other slowly varying
solar features. The temporal evolution of these filtered velocity signals for three
hours spanning the flare is shown in the upper panel of Figure~1(b) and that for
the quiet period is shown in Figure~2(b).

\subsection{WAVELET ANALYSIS OF VELOCITY DATA}

In order to examine the influence of flares on the high-frequency acoustic modes with time, we have
applied the Wavelet transform \citep{torrence98} on the velocity time series obtained from MDI
and GOLF as described in the previous Sections. We have used the Morlet wave as mother wavelet. 
Wavelet Power Spectrum (WPS) is computed, which yields the correlation between
the wavelet with a given frequency and the data along time. We limit our study to the region
inside a ``cone of influence'' corresponding to the periods of less than 25\% of the time series
length for reliability of the periods (c.f., \cite{mathur2010}). Finally, we have also calculated 
two confidence levels of detection corresponding to the
probability of 90\% and 50\% that the power is not due to noise. Thus, we have outlined the regions in the
WPS where power lies above these confidence levels and these regions are shown in middle panels of the
Figures~1-2(a)\&(b). A comparison of the WPS for the flare event with that of the quiet period
clearly indicates the enhancement of high-frequency waves during the flare as seen in the data
from MDI instrument. However, in the case of GOLF data, some short-lived high-frequency waves are
sporadically observed during the flares.

The WPS is collapsed along time to obtain the Global-wavelet power spectrum (GWPS). If some power is
present during the whole length of our time series, it would be easily seen in the GWPS. This is
nearly similar
to the commonly used power spectral density. In the Figures~1-2(a)\&(b), the GWPS shows a strong
peak of the normal
$p$ modes which are well known to be existing all the time. However, if the power of high-frequency
waves increases only a few times along the three hours of the studied data this would not appear as a
strong peak in the GWPS. In case of MDI data, the GWPS for the flare event do show a bump
corresponding to the high-frequency waves (above 5~mHz). However, the GWPS estimated
from the MDI velocity data during a quiet period (Figure~2(a)) doesn't show any peak beyond 5~mHz.
This supports the idea that the increase of power in the high frequency regime of the GWPS is
indeed caused by the flare. In the case of GOLF data, we do not observe this
signature in the GWPS estimated for the flare event as the overall high-frequency signal is weak in
these observations.

To see when the high-frequency waves have an increased power during the flare, we have calculated the
scale-average time series in the frequency regime 5-8~mHz. Basically, it is a collapsogram of the WPS
along the frequency of the wavelet in the chosen range. For this quantity, we have calculated the
confidence level for a 50\% probability. These are shown in the lower panels of the Figures~1-2(a)\&(b).
Here, we observe peaks corresponding to the presence of high-frequency waves in the WPS. In general, the
MDI data show more closely spaced high-amplitude peaks as compared to the GOLF data for the flare event,
but still mainly around 50\% confidence level. A comparison
with the same analysis performed for a quiet period (non-flaring condition) neither shows a high-frequency
bump beyond 5~mHz in the GWPS (for MDI data) nor high-amplitude peaks in the scale-average time series
(for both, MDI and GOLF data). Thus, in spite of the small confidence
levels found during this analysis, it indicates a possible relationship between these excess
high-frequency power and the flares.

\subsection{FOURIER ANALYSIS OF VELOCITY DATA}

We have also estimated the Fourier Power Spectrum (FPS) from the velocity time series obtained by the
MDI and GOLF instruments for the aforementioned flare event and the quiet period. The FPS spectra obtained
from the MDI and GOLF data
are respectively shown in the left and right panels of Figure~3. The FPS shown in 
the Figure~3 depicts dominant power in the 3~mHz band which
is due to the normal $p$ modes. These FPS also show significant spikes in the higher frequency
band (above 5~mHz) as estimated from MDI data for the flare event. However, the GOLF data shows marginal
signature for the flare event. At the same time, these spikes
are very weak in the case of a quiet period, as seen in both the data sets 
(c.f., bottom panels of Figure~3).

The difference found between the measurements of the two instruments could be a direct consequence
of the different heights of the solar atmosphere sampled by each instrument (Ni~I v/s Sodium doublet).

\section{DISCUSSION AND CONCLUSIONS}

Earlier attempts to find a correlation between
the energy of these high-frequency oscillations and flares using disk-integrated velocity observations
of the Sun had remained inconclusive \citep{gavry99, chap04}. In our analysis, the
enhancement of high-frequency power is clearly seen in the
MDI velocity data (and a feeble enhancement is also seen in the GOLF velocity data) during the
major flare event of 2003 October 28. It is in good agreement with the flare related enhancements
reported by \cite{karoff08} in disk-integrated intensity oscillations as observed with VIRGO.
Although, in our results the flare induced enhancement signals are seen with a low probability 
(around 50\%),
this signature is larger than that seen in non-flaring condition for both the MDI and GOLF data.

It is evident from the various models proposed for the generation of high-frequency waves that 
the amount of energy that is stored in the high-frequency waves is extremely
low compared to the amount of energy stored in the normal $p$ modes which are powered by the strong
turbulence in the convection zone of the Sun. Therefore, it is believed that the flare energy will have
a larger relative effect at high frequency as the other sources of its excitation are much smaller. A 
more detailed discussion on this can be found in \cite{kumar2010}. 

These observations provide us the opportunity to further investigate the possible excitation of 
global high-frequency waves
by local tremors due to major solar flares. As a future study, we plan to correlate the epochs of
enhancement of high-frequency waves with episodes of flare energy release seen in hard X-ray 
observations of the Sun. 

{\bf Acknowledgments}

The use of data from the MDI and the GOLF instruments on board {\em SOHO} spacecraft is gratefully
acknowledged.
The {\em SOHO} is a joint mission under cooperative agreement between ESA and NASA. This work has
been partially supported by the CNES/GOLF grant at the Service d'Astrophysique (CEA/Saclay). 
We are thankful to Douglas Gough, John Leibacher, Frank Hill, P. Scherrer, Robertus Erdelyi,
H. M. Antia, A. Kosovichev and Christoffer Karoff for useful discussions related to this work.


\end{document}